# Time-Series Based Thermography on Concrete Block Void Detection

Chongsheng CHENG[1] and Zhigang SHEN[2]

[1] The Durham School of Architectural Engineering and Construction, University of Nebraska-Lincoln, 113 Nebraska Hall, Lincoln, NE 68588-0500; e-mail: cheng.chongsheng@huskers.unl.edu

[2] The Durham School of Architectural Engineering and Construction, University of Nebraska-Lincoln, 113 Nebraska Hall, Lincoln, NE 68588-0500; e-mail: shen@unl.edu

## ABSTRACT

Using thermography as a nondestructive method for subsurface detection of the concrete structure has been developed for decades. However, the performance of current practice is limited due to the heavy reliance on the environmental conditions as well as complex environmental noises. A non-time-series method suffers from the issue of solar radiation reflected by the target during heating stage, and issues of potential non-uniform heat distribution. These limitations are the major constraints of the traditional single thermal image method. Time series-based methods such as Fourier transform-based pulse phase thermography, principle component thermography, and high order statistics have been reported with robust results on surface reflective property difference and non-uniform heat distribution under the experimental setting. This paper aims to compare the performance of above methods to that of the conventional static thermal imaging method. The case used for the comparison is to detect voids in a hollow concrete block during the heating phase. The result was quantitatively evaluated by using Signal-to-Noise Ratio. Favorable performance was observed using time-series methods compared to the single image approach.

## INTRODUCTION

Using thermography as a Nondestructive method for subsurface defects (e.g. delamination, voids) detection of the concrete structure has been developed for decades. The real-world implementation could relate to bridge deck inspection for delamination and concrete masonry wall grouting inspection for void (Khan et al. 2015). However, the performance of current practice is limited due to the heavy reliance on the environmental conditions as well as the complex environmental noises. Solar radiation intensity, wind speed, air temperature, humidity, and moisture have been reported as environmental factors that influence the appropriate time window of a day for conducting the inspection (Washer et al. 2009; Watase et al. 2015). Besides, the surface texture of the concrete, non-uniform heat distribution, and surface color difference have been observed to distract the accuracy of the detection based on the single thermogram analysis used in current practice.

In the field of infrared physics, the time series based methodologies have been widely used for subsurface abnormality detection for composites (Ibarra-Castanedo et al. 2009). Methodologies such as Fourier transform based pulse phase thermography (PPT) and principle component thermography (PCT) have been reported to have good tolerance on surface texture difference and non-uniform distribution under the experimental setting (Arndt 2010). Also, high order statistics (HOS) such as skewness and kurtosis of a distribution have been identified better performance to non-uniform heating (Madruga et al. 2010). However, full investigation of above methods has not been conducted on-field under environmental variations. As a start point, this paper aims to investigate the performance of PPT, PCT and HOS under the exposure of surface reflection and non-uniform temperature distribution in the experimental setting.

**BACKGROUND**

The principle of applying the thermography for subsurface anomaly detection is based on the temperature gradient difference between solid and debonded areas (see Figure1). The near surface delamination or void are treated as air block that slows down and reflects the directional heat transfer between the top and bottom of a reinforced concrete slab. Thus, the average heat transfer rate is different along the vertical direction. As a result, the surface temperature for debonded area is higher than the solid area during the heating phase (Figure 1(a)) and lower during the cooling phase of a day (Figure 1(b)). From the perspective of field inspection, the temperature contrast, optimal observation time window, environmental distractive sources, and relationship between depth and size are essential for the application.

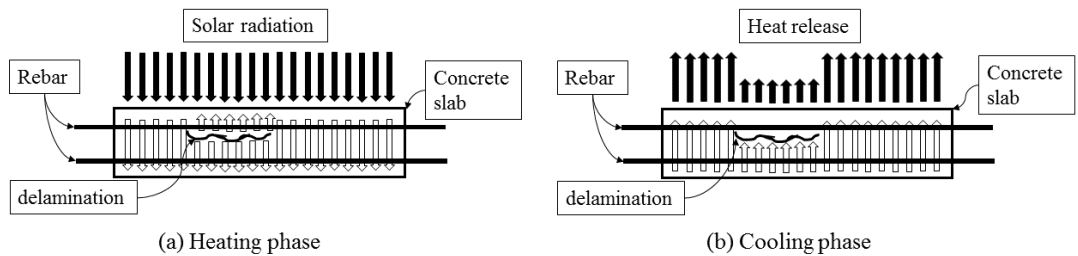

(a) Heating phase     (b) Cooling phase

**Figure1. Principle of detecting delamination at sub-surface of concrete slab during (a)heating and (b)cooling phases**

Temperature contrast is the fundamental criterion to distinguish the debonded area from the solid area. According to ASTM (2007), at least 0.5 °C difference between the debonded area and adjacent solid area was recommended. This recommendation has been considered suitable for daytime inspection with no shadowing effect and its feasibility was evaluated by Kee et al. (2011), and Omar and Nehdi (2016). In their studies, 7 hours after sunrise gave maximum contrast around 7 °C. In the condition that the structure was not under direct sunlight, Washer et al. (2013) reported that a minimum ambient temperature change of 1.5 Celsius degree per hour was necessary to have detectable contrast for shaded concrete structure for daytime inspection. However, other studies showed that the magnitude of temperature contrast was also related to the time window (Kee et al. (2011) and Watase et al. (2015)), depth difference and environmental conditions (Washer et al. 2009).

The favorable time window for data collection has been studied but contradictory outcomes were reported. Washer et al. (2009) recorded the occurrence time of maximum contrast for delamination in different depths and reported 5 to 9 hours after sunrise as the optimal observation time. Meanwhile, Kee et al. (2011) found 0.75 hours and 7.75 hours after sunrise were more favorable time window than 3.75 hours after sunrise. However, Hiasa et al. (2014) reported that the temperature data intends to be noisier in noon than in the evening due to possible influence from sun loading and surface texture of concrete. Thus, it recommended a night time as the prime data collection period even with the temperature contrast lower than 0.5 Celsius degree. The potential reason for those unclear might be the limited field experiments as well as variations in different climate regions and geographic differences.

Environmental factors and depth-to-size ratio has also been found to be related to the detectability. In ASTM (2007), environmental factors such as wind speed, sky clearness, and air temperature have been suggested for inspection. In Washer et al. (2009), a positive linear relationship was found between solar loading and temperature contrast. It also found the deeper defect had maximum temperature contrast at later time of the day. In their later research (Rumbayan and Washer 2014), solar radiation intensity, air temperature, and wind speed were included into a finite element model (FEM) for predicting the temperature contrast of debonded area to solid area in a concrete slab. Instead of the prediction based on the FEM, Watase et al. (2015) conducted a multi-regression model to predict the surface temperature for debonded and solid areas based on six parameters. Their analysis showed the ambient air temperature, temperature forecast, and pressure readings were the significant predictors. Although those environmental factors have been investigated, a cause-and-effect relationship was not yet clearly established. A systematic experimental study was conducted by Cotič et al. (2015) to determine the relationship between depth-to-size ratio and detectability. It was found that when the cover thickness was less than 0.9 times the size of the defects, the contrast method had the acceptable performance. However, this result was under a constant heating source in an experimental environment. Thus, there still has a gap for a practical evaluation.

The present implementation of thermography for anomaly detection within the concrete structure is limited by above factors due to the complexity of the natural environment and heat transfer behavior. Thus, the current contrast method based on the single thermal image may not be able to provide enough information to achieve the inspection requirement. Because of the variation in temperature contrast, time window and environmental conditions, the result suffered from errors and uncertainties which may lead to potential misjudgement. To address the problem, using time series data (sequential images) could be a promising solution to increase the detectability as well as the reliability. Related works are reviewed in the following section.

## RELATED WORKS

Several time series analyses for thermal images were reported useful to increase signal-to-noise ratio compared to the conventional contrast method. These methods were reported efficient in terms of the inhomogeneous sample surface and non-uniform heating under the experimental investigation. But there were limited studies conducted to apply them for the on-field application. Arndt (2010) adapted the square pulse phase thermography (PPT) for detecting the delamination-like defects inside the concrete slab for qualitative and quantitative investigation. The depth information for defect could be revealed by the phase image in the frequency domain. Dumoulin et al. (2013) developed a thermal imaging system for bridge inner structure condition monitoring. The system revealed the inner structures (e.g. beam and girders) layouts underneath the deck by using PPT and PCT analysis. Van Leeuwen et al. (2011) studied the PPT for honeycombing defects detection in concrete structures both under artificial heating and solar heating conditions. As far, applying time series thermographic analysis for inner defects detection of the concrete structure remains in the qualitative stage and the quantitative characterization of the defect was limited to the experimental condition with artificial heating.

From the perspective of inspection, the testing method should not only be robust to the environmental variations but also characterize the defect to support the engineering judgement. However, current literatures show both limitations in qualitative detection and quantitative characterization. Also, current time series analysis was conducted in the cooling phase only due to the high heating power during the heating phase that made the image saturated; and a pulse excitation required for depth estimation based on the simplified solution of 1D heating conduction equation (Arndt 2010). However, this will not be the case for the practical inspection that (1) natural solar radiation is not that intense; (2) the rough concrete surface scatters the radiation; (3) large thermal mass of concrete needs longer time for heating instead of a pulse excitation; (4) a heating stage lasts a long time during a day when taking the field inspection. Thus, it is worthwhile to investigate the time series analysis for the heating stage in terms of the detectability.

## METHODOLOGY

### Fourier Transform based PPT

Fourier transformation is a mathematic process that decomposes signal in the time order into a frequency order (see Figure2 (b)). By projecting in a way of sinusoids, the properties of the signal could be represented in terms of amplitude ($A_n$) and phase ($\emptyset_n$) in the frequency domain. The discrete Fourier transform can be written as follows:

$$F_n = \Delta t \sum_{k=0}^{N-1} T(k\Delta t) e^{-i2\pi nk/N} = Re_n + Im_n \quad (1)$$

$$A_n = \sqrt{Re_n^2 + Im_n^2} \text{ and } \emptyset_n = \tan^{-1}(Re_n/Im_n) \quad (2)$$

$$f_n = \frac{n}{N * \Delta t} \text{ and } \Delta f = \frac{\sum \Delta t}{N}$$

where $i^2 = -1$ is the imaginary number, $\Delta t$ is the sampling interval, $n$ is the integral increment in frequency ($n = 0, 1, \ldots, N$), N is the total number of frames that $N * \Delta t =$ duration of sampling, $T()$ is the temperature for each pixel at each frame. Then for the transform function $F$, at each frequency $F_n$, for a single pixel that is the weighted sum of all temperatures for that pixel across all time projecting into a sinusoid function. $Re_n$ and $Im_n$ are the real and imaginary part respectively. Solving the function can be achieved by using Fast Fourier Transform (FFT) in MATLAB.

Reconstruction of $A_n$ and $\emptyset_n$ will have the property image at the corresponding frequency. Especially the phase image is reported less affected by environmental reflections, emissivity variations, nonuniform heating, and surface geometry (Ibarra-Castanedo et al. 2009).

**Principle Component Thermography (PCT)**
Principle component thermography is an alternative transformation based tool to convert the data from spatial-temporal space into variation-in-orthonormal space (see Figure2 (c)). Singular Value Decomposition (SVD) is used to achieve the purpose. Since the SVD is based on matrix operation, the original data in the format of $x \times y \times N$ as a 3D matrix needs to be rearranged to $M \times N$ as a 2D matrix. Then the SVD of the 2D matrix can be solved by:

$$\boldsymbol{A = URV^T} \quad (3)$$

where **A** is the roaster matrix ($M \times N$) that $M = x * y$ and $N = frames$. **U** is the $M \times N$ orthogonal matrix, R is the $N \times N$ diagonal matrix storing the singular values, and $\boldsymbol{V^T}$ is the transpose of the $N \times N$ orthogonal matrix that characterizes the time variation in the order as principle components. Matrix **U** is the interest that describes the spatial variation in the orthogonal way which is referred as the empirical orthogonal functions (EOF). By remapping the each EOF in the column of **U** to a 2D matrix, an image can be reconstructed to distinguish defected and non-defected areas. (Ibarra-Castanedo et al. 2009)

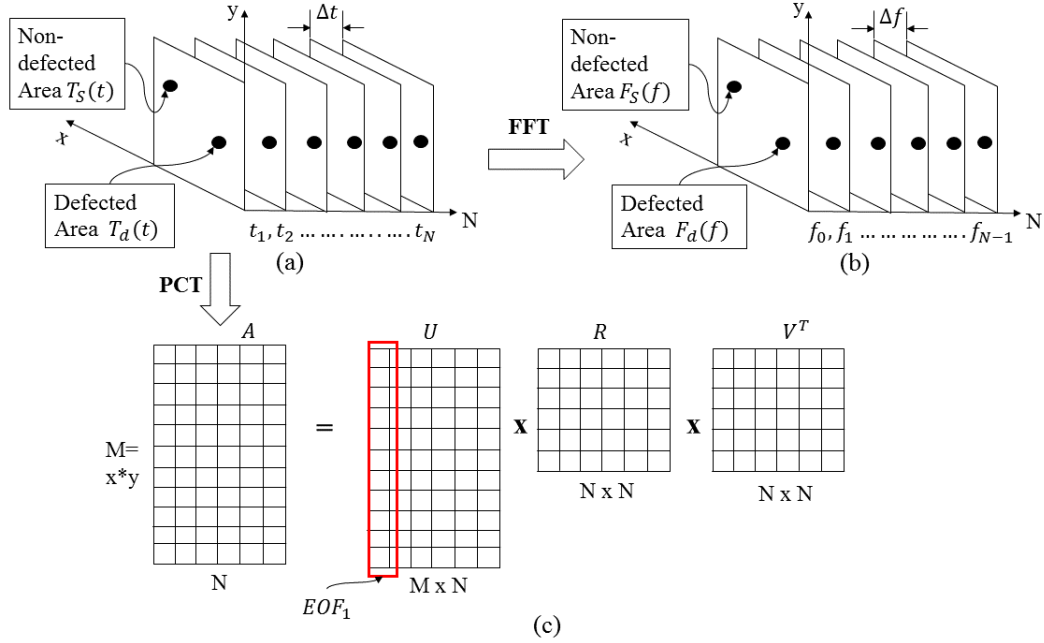

**Figure2. Time series thermal data structure and analysis. (a) raw temperature data; (b) Fourier Transform based PPT; (c) Principle Component Thermography**

**High-Order Statistics (HOS)**

High-order statistics such skewness and kurtosis are the commonly used statistical parameters to describe a distribution. It is reported that those parameters showed the high contrast in discriminating defected area under active thermography (Madruga et al. 2010). Skewness and kurtosis can be calculated by follows:

$$skewness = \frac{E[(X-\mu)^3]}{\sigma^3} \quad (4)$$

$$kurtosis = \frac{E[(X-\mu)^4]}{\sigma^4} \quad (5)$$

where $E[\,]$ is the expectation, μ is the mean, and $\sigma$ is the standard deviation. Skewness measures the asymmetry of the distribution and kurtosis describes the relative flatness of a distribution comparing to the normal distribution. In Madruga et al. (2010), the distribution of non-defected area presented higher skewness and kurtosis values than defected area.

**EXPERIMENT SETUP**

The experiment is set up (see Figure3 (b)) aiming to investigate the temperature evolution of hollow concrete block under artificial heating. It consists of test samples (laid concrete blocks), data collection equipment (Infrared camera) and heating source (halogen lamp). The sample layout is shown in Figure 3 (a), in which the interested hollow core concrete block is a CMU block generally used for foundation and masonry wall. The Infrared camera used here is FLIR A8300 which senses $3\sim5\mu m$ wavelength of infrared light and with maximum spatial resolution of 1280x720 pixels and thermal sensitivity of 0.02mK. Heating source is the halogen lamp with 4 bulbs summing up 1200W power. Data is recorded for 30 minutes with lamp heating at 0.2 Hz sampling rate and overall 360 frames of thermal images are collected.

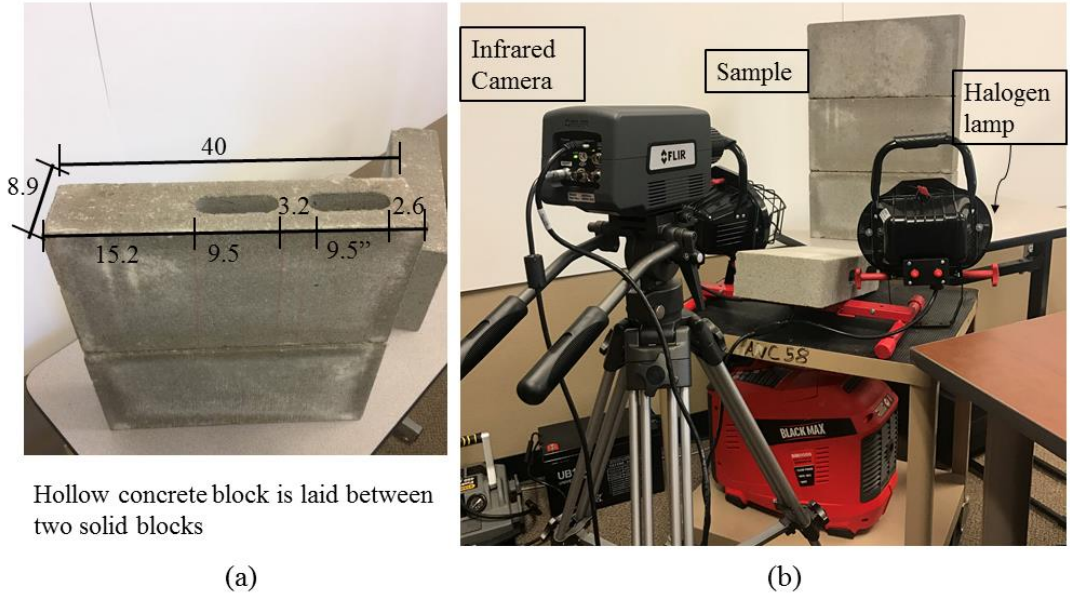

(a)                              (b)

**Figure3. Experiment setup. (a) sample dimension (cm); (b) experiment layout**

## DATA PROCESSING AND ANALYSIS

Collected thermal images are raw data of temperature values and thus several preprocessing are needed for later analysis. The data is stored in a 3D matrix that can be called $p(x, y, z)$ referring to the pixel value in the location of *(x,y)* of a thermal image at frame *z*. Then a spatial reduction is conducted to average out spatial noise as well as decreasing the requirement of computational power. Thus, the size of data reduces from 360x640x360 to 90x160x360 when using a 4x4 window to average. Under this processing, the size of each pixel is equivalents to $40(cm)/138(pixels) = 0.29(cm\ per\ pixel\ horizontal)$ which is still sufficient for the spatial resolution.

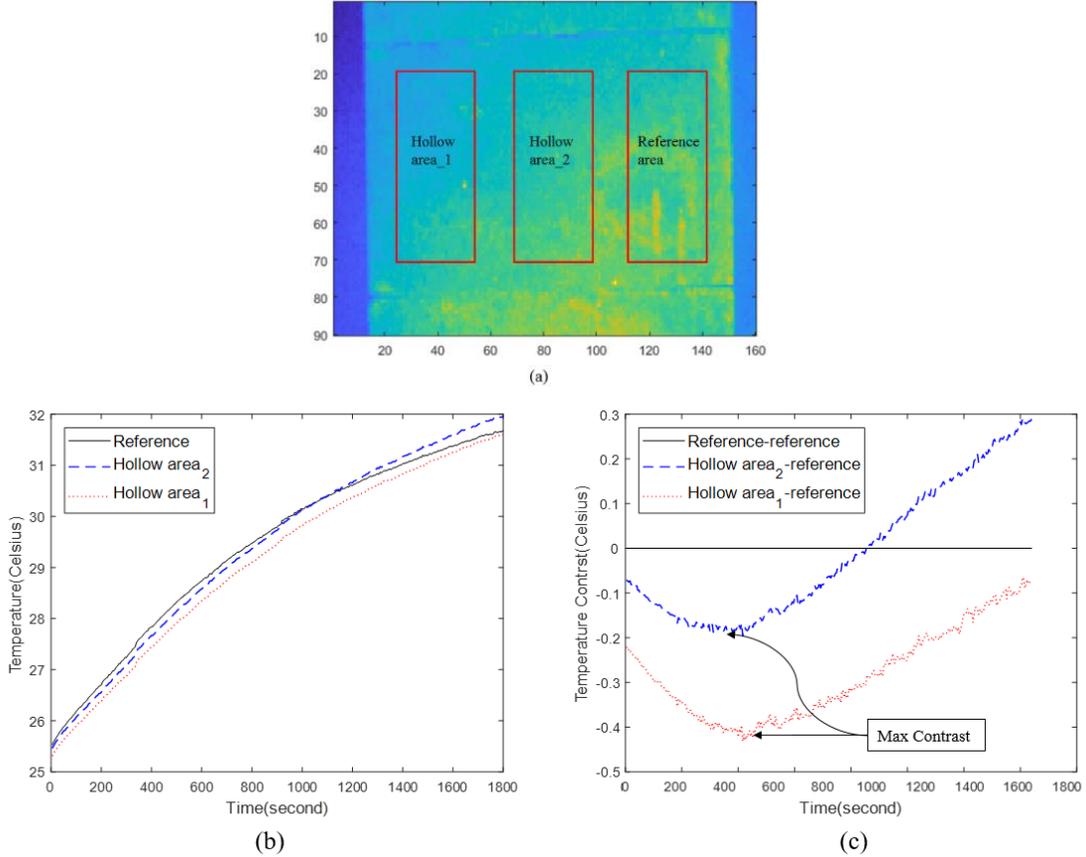

**Figure4. (a) Interested areas; (b) average temperature for each area; (c) average temperature contrast**

Interested areas are thereafter identified for quantitative comparison. Three areas are selected (see Figure 4(a)): reference area (area with solid concrete underneath), and hollow areas (area with hollows underneath). By plotting the mean of each area (see Figure 4(b)), the evolution behavior of reference is observed to be different from hollow areas; by plotting the contrast (hollow area's means − reference area's mean), we could find the roughly maximum contrast occurred around at 500 second as well as a little shift in time between two hollow areas (see Figure 4(c)). This shift indicates the non-uniformly distributed temperature due to the heating source and asymmetrical inner structure of the block. To this end, the question of detectability can be formulated to compare the contrast between the conventional single image contrast method and time series analysis.

To quantitatively evaluate the performance, the general measurements of Signal-to-Noise Ratio (SNR) is used. The calculation of SNR is as follows (Usamentiaga et al. 2014):

$$SNR = 20 log_{10}(\frac{|Def_\mu - Ref_\mu|}{Ref_\sigma})$$

where $Def_\mu$ is the mean of all pixels inside hollow area, $Ref_\mu$ is the mean of all pixels inside reference area, and $Ref_\sigma$ is the standard deviation of pixels in reference. Since the listed SNR takes the logomachic base of 10, the ratio in the parenthesis less than 10 will result in negative value in SNR. Negative SNR indicates the lack of detecting power. On the other hand, higher positive value presents high SNR referring higher contrast between reference and hollow areas in this case. To compare different time series methods consistently, the normalization of data is conducted to rearrange all pixel values of each processed image within [0,1]. The comparisons are shown in Table 1.

**Table 1. SNR Comparisons**

| Area | SNR (dB) | | | | | |
|---|---|---|---|---|---|---|
| | RAW | PCT | PPT-Phase | PPT-Amplitude | HOS-Skewness | HOS-Kurtosis |
| Hollow_1 | 1.38 | 21.63 | 18.83 | -1.60 | 11.06 | 14.51 |
| Hollow_2 | -2.93 | 20.75 | 18.68 | 0.41 | 11.63 | 14.76 |

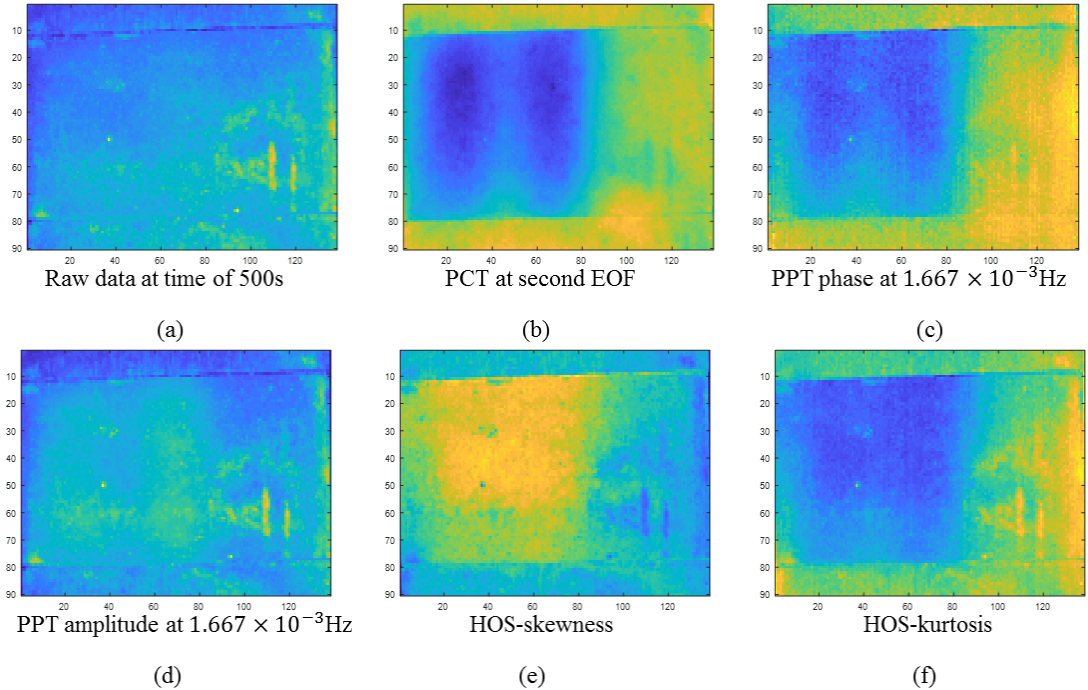

**Figure5. (a) raw image at 500 seconds heating; (b) PCT at 2nd EOF; (c) phase image of PPT; (d) amplitude image of PPT; (e) skewness image of HOS; (f) kurtosis image of HOS**

## DISCUSSION AND CONCLUSION

By comparing the SNR, a significant improvement can be found when using the time series analysis. According to Table 1, the PCT, phase image of PPT, and HOS increase the SNR from 1.38 (-2.93) up to 21.63 (20.75) correspondingly. Additionally, the above methods provide the closer value for each hollow area in SNR (21.63 for

hollow_1 and 20.75 for hollow_2) compare to raw data (1.38 and -2.93), which indicate these methods give the close measurements for the two hollowed areas. Since each hollowed area has the same depth from the surface with the major direction of heating conduction from front to bottom, the thermal response measured is expected to be the same for both hollow areas. Based on this assumption, the time series analysis provide more reliable metrics than contrast method in terms of non-uniformed heating condition.